\documentclass[prb,twocolumn,showpacs,floatfix,preprintnumbers,amsmath,amssymb,superscriptaddress]{revtex4-1}
\usepackage{graphicx}
\usepackage{dcolumn}
\usepackage{bm}
\usepackage{color}
\usepackage[latin1]{inputenc}
\usepackage[urlcolor=blue]{hyperref}
\hypersetup{backref, colorlinks=true}

\usepackage{color}
\usepackage[latin1]{inputenc}

\begin{document}

\preprint{APS/123-QED}

\title{Pr magnetism and its interplay with the Fe spin density wave in PrFeAsO}

\author{U.~Stockert}\email{ulrike.stockert@cpfs.mpg.de}\affiliation{MPI for Chemical Physics of Solids, D-1187 Dresden, Germany}\affiliation{Institute for Solid State Research, IFW Dresden, D-01171 Dresden, Germany}
\author{N.~ Leps}\affiliation{Institute for Solid State Research, IFW Dresden, D-01171 Dresden, Germany}\affiliation{Kirchhoff Institute for Physics, D-69120 Heidelberg, Germany}
\author{L.~Wang}\affiliation{Institute for Solid State Research, IFW Dresden, D-01171 Dresden, Germany}\affiliation{National High Magnetic Field Laboratory, Florida State University, Tallahassee, Florida 32310-3706, USA}
\author{G.~Behr}\affiliation{Institute for Solid State Research, IFW Dresden, D-01171 Dresden, Germany}
\author{S.~Wurmehl}\affiliation{Institute for Solid State Research, IFW Dresden, D-01171 Dresden, Germany}
\author{B.~B\"uchner}\affiliation{Institute for Solid State Research, IFW Dresden, D-01171 Dresden, Germany}\affiliation{Institute for Solid State Physics, TU Dresden, D-01069 Dresden, Germany}
\author{R.~Klingeler}\affiliation{Kirchhoff Institute for Physics, D-69120 Heidelberg, Germany}

\date{\today}

\begin{abstract}
We have studied the magnetism of the Pr$^{3+}$ ions in
PrFeAsO$_{1-x}$F$_x$ ($x = 0, 0.15$) and its interaction with the Fe
magnetic order (for $x = 0$). Specific heat data confirm the
presence of a first excited crystal electric field (CEF) level
around 3.5~meV in the undoped compound PrFeAsO. This finding is in
agreement with recent neutron scattering experiments. The doped
compound is found to have a much lower first CEF splitting of about
2.0 meV. The Pr ordering in PrFeAsO gives rise to large anomalies in
the specific heat and the thermal expansion coefficient. In
addition, a field-induced transition is found at low temperatures
that is most pronounced for the magnetostriction coefficient. This
transition, which is absent in the doped compound, is attributed to
a reversal of the Fe spin canting as the antiferromagnetic Pr order
is destroyed by the external magnetic field.
\end{abstract}


\maketitle

\subsection{Introduction}

Layered FeAs-materials have been studied extensively since the
discovery of superconductivity in LaFeAsO$_{1-x}$F$_x$ with
transition temperatures $T_\mathrm{c}$ up to
28~K.~\cite{Kamihara2008} By exchanging the nonmagnetic La with
magnetic rare earths (RE) such as Pr or Sm the superconducting
transition temperature could be increased above
50~K.~\cite{Ren2008b,Liu2008a} All parent compounds of the
\mbox{\textit{RE}FeAsO} family (\textit{RE} = La, Ce, Pr, Nd, Sm,
Gd) behave rather similar:~\cite{F4-08-12, F4-08-10, F4-08-26,
F4-08-25, F4-08-15, F4-08-17, F4-10-2, Alfonsov2011} They exhibit
magnetic ground states with a structural distortion from a
tetragonal to an orthorhombic lattice around 150~K and subsequent
formation of a spin density wave (SDW) of the Fe moments. Except for
LaFeAsO additional antiferromagnetic (AFM) ordering of the RE
moments is observed with transition temperatures of the order of
10~K.

PrFeAsO has the highest RE ordering temperature among the
\textit{RE}FeAsO family with $T_\mathrm{N}^{\mathrm{Pr}} \approx
12$~K.~\cite{F4-08-11,F4-08-10} The structural and SDW transitions
take place around 153~K and 127~K, respectively.~\cite{F4-08-10} A
delicate interaction between the Fe and Pr moments has been deduced
from $\mu$SR~\cite{F4-09-5} and M\"{o}ssbauer~\cite{F4-09-3}
experiments such that the ordering of each sublattice entails a
reorientation of the other one. The magnetic properties of PrFeAsO
are further complicated by crystal electric field (CEF) effects: In
tetragonal symmetry the free-ion ground-state multiplet of Pr$^{3+}$
is split into 5 singlets and 2 doublets, while the remaining
degeneracy is lifted in any lower symmetry.~\cite{G-84-1} Inelastic
neutron scattering (INS) experiments revealed a first CEF excitation
around 3.5~meV that is split above
$T_\mathrm{N}^{\mathrm{Pr}}$.~\cite{F4-11-1} However, the energetic
position of the other levels remained unclear. Substitution of O by
F changes the magnetism fundamentally: It leads to a simultaneous
destruction of the Pr and Fe order.~\cite{F4-09-4} Moreover,
additional CEF excitations have been found around 10~meV for $x =
0.13$.~\cite{F4-11-1}

In this paper we investigate the Pr magnetism in PrFeAsO and its
interplay with the Fe SDW. We start with an estimation of the CEF
level scheme of Pr$^{3+}$ in PrFeAsO$_{1-x}$F$_x$ ($x = 0, 0.15$).
Fluorine doping is found to have a significant influence on the
splitting, in agreement with the INS results. We find an additional
low-lying level around 2~meV in the F-doped compound, which was not
seen before. The sensitive dependence of the splitting on the F
substitution is discussed a possible reason for the absence of Pr
magnetic order in the doped compound. Subsequently we present
thermal expansion, magnetostriction and magnetization data. PrFeAsO
is found to undergo a field-induced transition below
$T_\mathrm{N}^{\mathrm{Pr}}$, which is attributed to an Fe spin
reorientation due to the destabilization of the AFM Pr order.


\subsection{Experimental details}

Polycrystalline samples have been prepared by solid state reaction
as described in Ref.~\onlinecite{Kondrat2009}. The specific heat was
studied in a Quantum Design PPMS by means of a relaxation technique.
For the thermal expansion and magnetostriction measurements a
capacitance dilatometer was utilized, which allows a very accurate
study of sample length changes $dL/L$.~\cite{Wang2009} The linear
thermal expansion coefficient $\alpha$ was calculated as the first
temperature derivative of $dL/L$, while the magnetostriction
coefficient $\beta$ is determined by the first derivative of $dL/L$
with respect to the magnetic field $B = \mu _0 H$. Magnetization
measurements were performed in a commercial VSM-Squid (Quantum
Design) at an oscillation frequency of 14~Hz.


\subsection{The CEF splitting of Pr$^{3+}$}

\begin{figure}[tb]
\begin{center}
\includegraphics[width=0.48\textwidth]{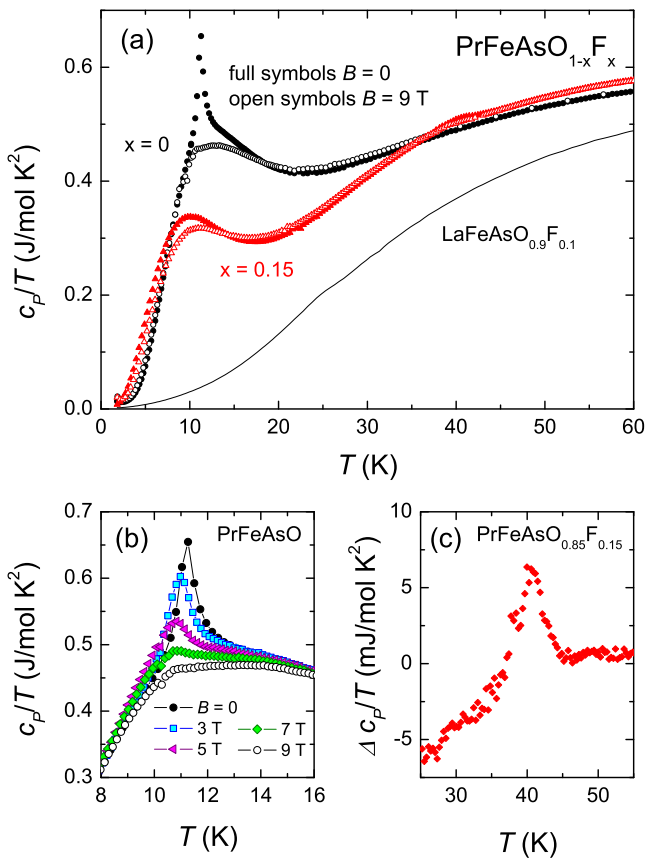}
\end{center}
\caption{(a) Specific heat $c_p$ divided by temperature $T$ for
PrFeAsO$_{1-x}$F$_x$ ($x = 0, 0.15$) in zero magnetic field and 9~T.
For $x = 0$ a clear anomaly with a maximum at 11.3~K is observed in
zero field originating from the magnetic ordering of the Pr$^{3+}$
ions at $T\mathrm{_N^{Pr}}$. The specific heat of
LaFeAsO$_{0.9}$F$_{0.1}$ used as a phonon background is also shown
for comparison. (b) Behavior of the anomaly at $T\mathrm{_N^{Pr}}$
of PrFeAsO in magnetic field. (c) The superconducting transition of
the sample with $x = 0.15$ gives rise to a small anomaly seen in
$\Delta c_p/T = c_p(0)/T-c_p(9\,\mathrm{T})/T$. \label{cPvsT}}
\end{figure}

First we will evaluate the crystal electric field (CEF) level scheme
of the Pr$^{3+}$ ions and the influence of fluorine doping by
analyzing the specific heat of both compounds.

Fig.~\ref{cPvsT}a shows the specific heat $c_p$ of
PrFeAsO$_{1-x}$F$_x$ ($x = 0, 0.15$) as $c_p/T$ for $T \leq 60$~K.
Data for \mbox{$x = 0$} at higher $T$ are similar to those published
in Ref.~\onlinecite{F4-11-2} with anomalies due to the structural
transition at $T_s = 145$~K and the SDW formation at
$T_\mathrm{N}^{\mathrm{Fe}} = 129$~K.~\cite{NLeps} These anomalies
are absent for $x = 0.15$. Here, we focus on the regime below 60~K.
The undoped sample exhibits a sharp anomaly with a maximum at
11.3~K, which is on top of a broad hump. This behavior is attributed
to the magnetic ordering of the Pr$^{3+}$ moments in the presence of
CEF splitting. The evolution of the anomaly in an applied magnetic
field is shown in Fig.~\ref{cPvsT}b. Small fields of 3~T and 5~T
lead to a weak shift to lower $T$ and a reduction in height. In 7~T,
the anomaly is hardly seen anymore. Application of a field of 9~T
suppresses the anomaly completely and leaves only the hump and a
tiny kink. This behavior is in line with a suppression of the
antiferromagnetic Pr order by the external magnetic field. On the
other hand, the hump due to the CEF splitting is scarcely changed by
the field, because the corresponding energy scale is much larger
than the Zeeman splitting. For the sample with 15~\% fluorine
doping, no Pr ordering is observed down to 1.5~K, and the hump is
shifted to lower $T$. In addition, a small anomaly is found at the
superconducting transition, which is largely suppressed in 9~T. The
difference between the zero-field and 9 T data reveals a small
$\lambda$-shaped anomaly at $T_\mathrm{c} = 42$~K
(cf.~Fig.~\ref{cPvsT}c).

\begin{figure}[tb]
\begin{center}
\includegraphics[width=0.48\textwidth]{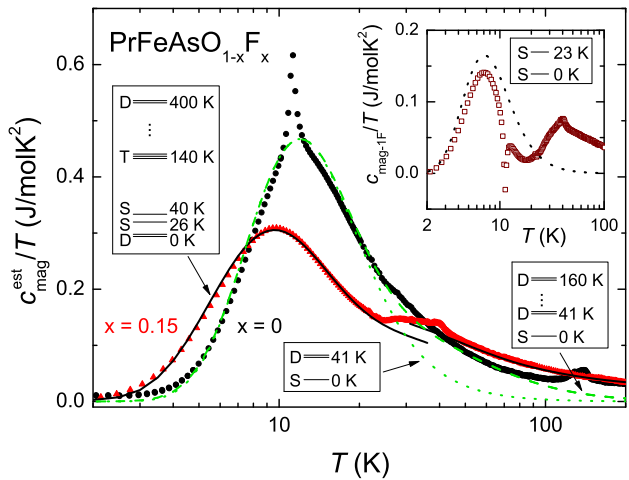}
\end{center}
\caption{Estimate of the magnetic contribution to the specific heat
of PrFeAsO$_{1-x}$F$_x$ ($x = 0, 0.15$) calculated by subtracting
the data of LaFeAsO$_{0.9}$F$_{0.1}$. The solid lines are the
results of a fit to a Schottky contribution from CEF splitting for
$x = 0.15$. Above $T_\mathrm{c}$ an additional charge carrier
contribution $\Delta \gamma = 25.2$~mJ/mol K is taken into account,
which leads to a shift of the line. The dotted line is the
calculated Schottky contribution for a ground state singlet and an
excited doublet at 41~K. The agreement with the data for $x  =0$ is
much improved by adding a second doublet at 160~K (dashed line). The
inset shows the estimated contribution to $c_{\mathrm{mag}}$ of $x =
0.15$ from sites with one or more fluorine neighbors in a model with
different CEF schemes for differing environments~\cite{F4-11-1} as
explained in the text. The dotted line is the calculated
contribution expected for two singlet states at 0 and 23~K for 37\%
of the Pr sites. \label{DeltacP}}
\end{figure}

There is a significant magnetic contribution $c_{\mathrm{mag}}$ to
the specific heat of PrFeAsO$_{1-x}$F$_x$ stemming from the
Pr$^{3+}$ ions. An exact determination of $c_{\mathrm{mag}}$ is
impossible due to the presence of the various phase transitions. In
order to get a rough estimate of the magnetic contributions
$c_{\mathrm{mag}}^\mathrm{{est}}$, we used LaFeAsO$_{0.9}$F$_{0.1}$
as a nonmagnetic reference. This is reasonable as substitution of
the rare-earth ion is not expected to change the phonon spectrum
significantly and the anomaly due to the superconducting transition
of LaFeAsO$_{0.9}$F$_{0.1}$ at 26 K is very small. Therefore, its
contribution to $c_p$ is negligible compared to the Pr contribution.
However, one has to keep in mind, that there might be a different
electronic contribution to $c_p$ (see discussion below).
$c_{\mathrm{mag}}^\mathrm{{est}}$ is obtained by subtracting the
specific heat of LaFeAsO$_{0.9}$F$_{0.1}$, which is shown for
comparison in Fig.~\ref{cPvsT}a, from the one of
PrFeAsO$_{1-x}$F$_x$. The result is plotted in Fig.~\ref{DeltacP} as
$c_{\mathrm{mag}}^\mathrm{{est}}/T$. Both PrFeAsO$_{1-x}$F$_x$
samples exhibit Schottky-like anomalies, whereas the maximum occurs
at lower $T$ for the sample with fluorine doping. This demonstrates
already that the doping changes the CEF level scheme and results in
a lowering of the first excited level.

For PrFeAsO$_{0.85}$F$_{0.15}$ no magnetic ordering is observed down
to 1.5 K and $c_{\mathrm{mag}}$ is only due to the thermal
occupation of higher CEF states. We may use
$c_{\mathrm{mag}}^\mathrm{{est}}$ to estimate the CEF splitting. The
data range between 20~K and 45~K, which contains the superconducting
anomalies, was omitted for the fitting. The best description of the
data is obtained for a ground-state doublet (D) with excited
singlets (S) at 26~K and 40~K. Two more levels are found at 140~K
(quasi-triplet (T)) and at 400~K (quasi-doublet). These two levels
may in fact consist of three or two close-lying singlets, which,
however, cannot be resolved from our data. In our fit we also
accounted for a difference between the electronic contributions of
LaFeAsO$_{0.9}$F$_{0.1}$ and PrFeAsO$_{0.85}$F$_{0.15}$. Below the
superconducting transitions it was taken zero assuming that all
charge carriers are condensed to Cooper pairs. Above $T_\mathrm{c}$
we obtain a value of $\Delta \gamma =
25$~mJ/$(\mathrm{mol}\,\mathrm{K}^2)$ in addition to the Sommerfeld
coefficient of LaFeAsO$_{0.9}$F$_{0.1}$ of
3.5~mJ/$(\mathrm{mol}\,\mathrm{K}^2)$.~\cite{NLeps} Thus, we may
estimate $\gamma = 31.5$~mJ/$(\mathrm{mol}\,\mathrm{K}^2)$ for
PrFeAsO$_{0.85}$F$_{0.15}$, which is comparable to the value of
34~mJ/$(\mathrm{mol}\,\mathrm{K}^2)$ determined for
PrFeAsO.~\cite{F4-11-2}

The result of the fitting is shown as lines in Fig.~\ref{DeltacP}
together with a schematic drawing of the corresponding level scheme.
Our fit describes the data very well. However, the proposed
ground-state doublet raises the question, why no magnetic ordering
is observed in PrFeAsO$_{0.85}$F$_{0.15}$. We would like to mention,
that a description of the data using a ground-state singlet was not
possible. This might indicate that $c_{\mathrm{mag}}^\mathrm{{est}}$
is not sufficiently precise to allow estimating the CEF
contribution. However, the presence of two singlets at 26~K and 40~K
above a ground-state doublet suggested by our fit is determined
mainly by the data below 20~K. In this temperature range, the
estimate for $c_{\mathrm{mag}}$ is rather good, because the phonon
contribution is comparably small. A possible answer is given by a
recent neutron diffraction experiment. Goremychkin \textit{et
al.}~studied CEF excitations in PrFeAsO$_{1-x}$F$_x$ ($x = 0,
0.13$).~\cite{F4-11-1} The undoped compound was found to have an
excitation around 3.5~meV that is split above
$T_\mathrm{N}^{\mathrm{Pr}}$. Substitution of fluorine removed the
splitting and led to a reduction of that peak. In addition, two more
peaks were observed at 9.7~meV and 11.8~meV. The authors explained
their observation by the presence of two different well-defined
charge environments resulting from a random distribution of fluorine
on the oxygen sites. This leads to 5 different nearest-neighbor
configurations for the Pr$^{3+}$ ions, the most common of which are
those with none (58~\%) and one (35~\%) F neighbor. The excitation
at 3.5~meV with reduced height was attributed to the Pr$^{3+}$ ions
with unchanged nearest-neighbor configuration, while the two
additional peaks were attributed to Pr$^{3+}$ ions with one fluorine
as nearest neighbor.

This idea can be also applied to our specific heat data: We assume
that $c_{\mathrm{mag}}^\mathrm{{est}}$ of PrFeAsO$_{0.85}$F$_{0.15}$
contains a CEF contribution from the 52~\%~Pr$^{3+}$ ions (for $x =
0.15$) without fluorine neighbor. It can be estimated from
$c_{\mathrm{mag}}^\mathrm{{est}}$ of PrFeAsO, because the entropy
change related to the magnetic ordering in the undoped compound is
rather small. After subtracting this contribution we end with a
magnetic specific heat $c_{\mathrm{mag-1F}}$ dominated by the CEF
contribution from the 37~\% ions with one F neighbor. The resulting
curve is shown in the inset of Fig.~\ref{DeltacP}. The
low-temperature part is dominated by a Schottky-like anomaly, which
is surprisingly well described by two singlets at 0 and 23~K at
37~\% of the Pr sites. Due to the various approximations used so
far, we refrain from a further analysis of $c_{\mathrm{mag-1F}}$.
However, the broad maximum observed at higher temperatures may be
ascribed to the CEF levels around 10~meV found in the neutron
scattering experiments. The splitting of 23~K determined from
$c_{\mathrm{mag-1F}}$ is close to the lowest splitting of 26~K
determined by our fit of $c_{\mathrm{mag}}^\mathrm{{est}}$.
Therefore, our data clearly prove the presence of a first excited
level at about 2~meV for PrFeAsO$_{0.85}$F$_{0.15}$. No respective
excitation was observed in the neutron scattering experiment. We
suggest, that it probably corresponds to a forbidden transition.

We now turn to the undoped compound. A CEF excitation around 3.5~meV
has been found in the neutron scattering experiment mentioned
above.~\cite{F4-11-1} We use this value to model our magnetic
specific heat. A quite good description of the low-temperature part
is indeed obtained assuming a singlet ground-state with a doublet at
41~K corresponding to 3.5~meV (cf. dotted green line in
Fig.~\ref{DeltacP}). The agreement at high $T$ is significantly
improved by assuming a second doublet at 160~K, which is at the
limit of the measurement range in Ref.~\onlinecite{F4-11-1}. The
other states are supposed to lie at even higher energy.

Summarizing the analysis presented so far we conclude the following:
Our specific heat data for PrFeAsO are consistent with a singlet
ground state and excited doublets around 41~K and 160~K, whereas the
remaining levels lie at higher energy. PrFeAsO$_{0.85}$F$_{0.15}$
has a first excited state at significantly lower energy
corresponding to about 23-26~K. This low-lying singlet is
responsible for the shift of the hump in $c_p$ to lower $T$
(cf.~Fig~\ref{cPvsT}a). Our data are also in line with neutron
scattering results, which suggest different Pr sites for
PrFeAsO$_{0.85}$F$_{0.15}$, depending on their environment. In this
model, 52~\% of the Pr ions have no fluorine neighbor and CEF levels
similar to PrFeAsO with a first excited state being a doublet at
41~K. The 37~\% Pr ions with one fluorine neighbor have a different
splitting, whereas the first excited state is a singlet at 23~K.

The change of the CEF level scheme upon F substitution is probably
also responsible for the absence of the Pr magnetic order in
PrFeAsO$_{0.85}$F$_{0.15}$. Another possible reason might be the
absence of the Fe SDW. In fact, it has been found that both F doping
and Ru substitution on the Fe site lead to a concomitant suppression
of $T_\mathrm{N}^{\mathrm{Fe}}$ and
$T\mathrm{_N^{Pr}}$.~\cite{F4-09-4, F4-12-2} This suggests that both
orderings are linked. However, competing magnetic structures have
been proposed for the Fe and Pr sublattices,~\cite{F4-09-5} as
explained in more detail below. This renders such a connection
rather unlikely. Instead, in view of the sensitive dependence of the
first excited CEF level on the Pr environment we suggest that Pr
ordering takes place only for a specific CEF splitting. One may even
speculate that the "CEF disorder" due to the presence of different
Pr environments and consequently different CEF states is responsible
for the absence of Pr magnetic order in PrFeAsO$_{0.85}$F$_{0.15}$.

\subsection{The interplay of Fe and Pr magnetism}

\begin{figure}[tb]
\begin{center}
\includegraphics[width=0.48\textwidth]{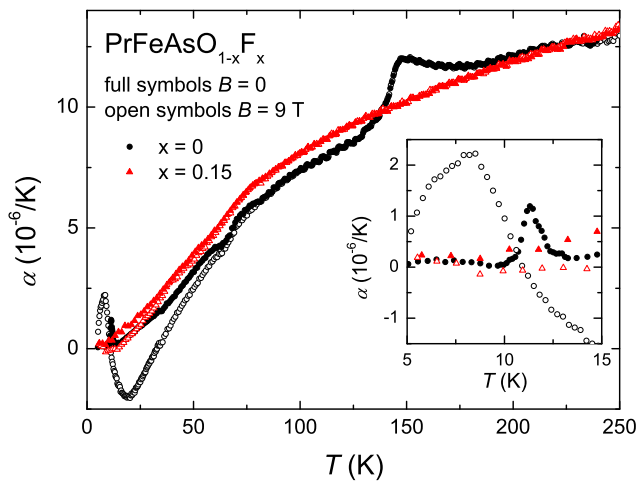}
\end{center}
\caption{Temperature dependence of the linear thermal expansion
coefficient $\alpha (T)$ of PrFeAsO$_{1-x}$F$_x$ ($x = 0, 0.15$) in
magnetic fields of 0~T and 9~T. Clear anomalies are seen in the
zero-field thermal expansion curve for $x = 0$ at the structural/SDW
transitions and at the magnetic ordering temperature of the
Pr$^{3+}$ ions. These are absent for the doped compound. Application
of a magnetic field leads to a strong change of $\alpha (T)$ for
PrFeAsO below 70~K. By contrast, there is almost no effect on
$\alpha (T)$ for the doped compound. The inset shows the low-$T$
part on a larger scale. \label{texp}}
\end{figure}

We now turn to the interplay of Fe and Pr magnetism in PrFeAsO. For
this purpose we present thermal expansion, magnetostriction, and
magnetization data.

The linear thermal expansion coefficient $\alpha$ of
PrFeAsO$_{1-x}$F$_x$ ($x = 0, 0.15$) measured in zero magnetic field
and 9~T is shown in Fig.~\ref{texp}. The structural and SDW
transitions of the undoped compound give rise to a large, broad
anomaly around 150~K.~\cite{Klingeler2009-ICM} Both transitions
cannot be distinguished in our data, probably due to the small
height of one of the anomalies.
The ordering of the Pr$^{3+}$ moments in PrFeAsO shows up as
another, positive anomaly with a maximum at 11.4~K, close to the
maximum in $c_p /T$. This anomaly is rather sharp and symmetric,
which suggests that the transition might in fact be of first-order
type. The shape of our specific-heat anomaly is also in line with a
broadened first-order transition, yet we cannot exclude, that it is
a second-order one.

Although the pronounced background hinders the precise determination
of the anomalous changes, the hydrostatic pressure dependence can be
extracted quantitatively from the anomalies in $\alpha$ and $c_p$ by
means of either the Clausius-Clapeyron or the Ehrenfest relation
depending on the nature of the phase transition. Supposing a weak
first-order character of the anomaly, analyzing the data in
Fig.~\ref{cPvsT}b and \ref{texp} yields a volume change of $\Delta V
= \int 3\cdot \Delta\alpha(T)dT \approx 3.9\cdot 10^{-6}$/K and an
entropy change of $\Delta S \approx 0.13$~J/(mol K) at the
transition. Applying these estimates we obtain the hydrostatic
pressure dependence:
\begin{equation}
\frac{\partial T\mathrm{_N^{Pr}}}{\partial p}=V_{\rm mol}\cdot
\frac{\Delta V}{\Delta S}\approx 1.2\, {\rm K/GPa}.
\end{equation}
A similar value is obtained assuming that the transition is of
second order. In this case the pressure dependence is determined
from the jump heights $\Delta \alpha \approx 1.0\cdot 10^{-6}$/K and
$\Delta c_p \approx 1.4$~J/(mol K), which yields:
\begin{equation}
\frac{\partial T\mathrm{_N^{Pr}}}{\partial p}=T V_{\rm mol}\cdot
\frac{3 \Delta \alpha}{\Delta c_p}\approx 1.0\, {\rm K/GPa}.
\end{equation}
So far, no pressure experiments have been performed on PrFeAsO.
However, the estimated pressure dependence of $T\mathrm{_N^{Pr}}$ is
comparable to the one of the Ce ordering temperature in CeFeAsO of
0.9 GPa/K.~\cite{F4-11-7}

A field of 9~T leads to a strong decrease of the thermal expansion
coefficient of PrFeAsO below approximately 50~K.  The sharp anomaly
at 11~K is suppressed. Instead a broad maximum is observed around
8~K as seen in the inset of Fig.~\ref{texp}, which shows the low-$T$
part of $\alpha (T)$ on a larger scale. Since no anomaly is found in
the corresponding specific heat curve, this feature is not related
to a phase transition, but rather to thermal population of higher
Pr$^{3+}$ states. Contrary to the undoped compound, the thermal
expansion coefficient of PrFeAsO$_{0.85}$F$_{0.15}$ exhibits a
smooth temperature dependence. No anomaly is found at the
superconducting transition similar to the findings in
LaFeAsO$_{1-x}$F$_x$.~\cite{Wang2009} Most probably, it is too small
to be seen in our data. Application of a magnetic field of 9~T leads
to a lowering of $\alpha$, which, however, is much weaker than for
the undoped system.

\begin{figure}[tb]
\begin{center}
\includegraphics[width=0.48\textwidth]{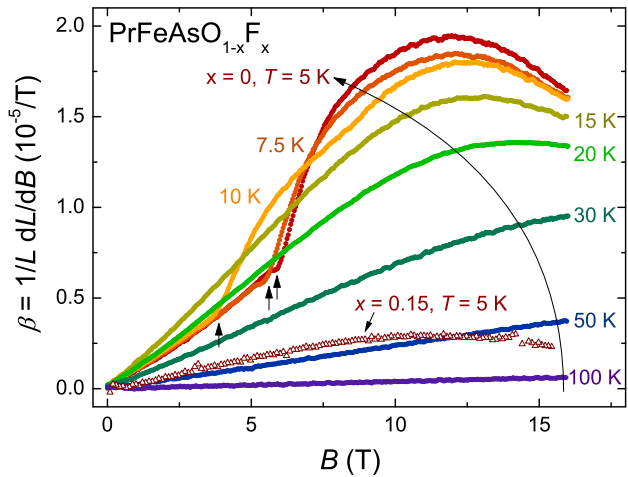}
\end{center}
\caption{Magnetostriction coefficient $\beta$ vs.~magnetic field $B$
of PrFeAsO at different temperatures up to 100~K (full symbols). For
$T = 5$~K, 7.5~K, and 10~K kinks marked by arrows are observed at
5.8~T, 5.4~T, and 3.7~T, respectively. At higher $T$ the transition
disappears. For comparison data for PrFeAsO$_{0.85}$F$_{0.15}$
measured at 5~K are shown (open symbols). At higher temperature,
$\beta$ is even smaller for this compound. \label{ms}}
\end{figure}

In view of the large change of $\alpha$ of PrFeAsO induced by a
magnetic field we expect a sizeable magnetostrictive effect in the
material. In Fig.~\ref{ms} we present the magnetostriction
coefficient $\beta =1/L \, \mathrm{d} L/\mathrm{d}B$ of PrFeAsO at
different temperatures up to 100~K and in magnetic fields up to
16~T. At high temperatures, an almost linear dependence $\beta(B)$
is found. With decreasing $T$, the magnetostriction coefficient
increases, and a maximum in $\beta(B)$ develops at higher fields.
Below the Pr ordering temperature, the behavior of $\beta(B)$
changes qualitatively. In low fields, $\beta(B)$ increases linearly
with a temperature-independent slope. Interestingly, this slope is
smaller, than the one observed at 15~K, i.e.~above the phase
transition. At higher fields, kinks are observed in $\beta(B)$ at
5.8~T, 5.4~T, and 3.7~T for 5~K, 7.5~K and 10~K, respectively. For
fields larger than approximately 10~T, the low-$T$ magnetostriction
resembles the field dependence observed above 15~K. For comparison,
a curve measured at 5~K on PrFeAsO$_{0.85}$F$_{0.15}$ is also shown
in Fig.~\ref{ms}. As expected from the small change of the thermal
expansion coefficient in magnetic field, the magnetostriction is
much weaker than for the undoped compound.

The magnetostrictive effect observed in PrFeAsO below about 50 K,
i.~e.~well above $T_\mathrm{N}^{\mathrm{Pr}}$ is rather large. This
is surprising since the magnetostrictive effect for
PrFeAsO$_{0.85}$F$_{0.15}$ is significantly weaker. Apart from the
Pr magnetic ordering there are two important differences between the
undoped and the doped compound: (1) In addition to the Pr order, the
Fe spins form a SDW below 129~K in PrFeAsO. It may not be directly
responsible for the large magnetostriction. For comparison:
measurements on LaFeAsO allow estimating an upper limit of $\beta <
5\times 10^{-8}/\mathrm{T}$ for this compound, despite the presence
of Fe SDW order.~\cite{LWang} However, the Fe SDW induces a
polarization of the Pr$^{3+}$ moments already well above
$T_\mathrm{N}^{\mathrm{Pr}}$.~\cite{F4-09-5} The applied magnetic
field thus acts on a compound with net magnetic moments from both
the Fe SDW and the Pr sublattice, which may be the reason for the
large magnetostrictive effect. (2) As evident from neutron
scattering data~\cite{F4-11-1} and our specific heat analysis,
fluorine substitution changes the CEF level scheme of part of the
ions. This may explain the observed suppression of $\beta(B)$ at
least to some extent. Only 52~\% Pr ions without fluorine neighbor
have an unchanged CEF scheme. Therefore, one expects about half the
magnetostrictive effect of PrFeAsO for the fluorine-doped compound
from these sites. Moreover, we do not know the magnetostriction
contribution of the remaining 48~\% Pr ions with at least one
fluorine neighbor. In addition, the strain induced by the differing
favored expansion coefficients of neighboring cells will lead to a
more difficult behavior than a simple sum of the effects.
Measurements on samples with different fluorine content might
clarify the relevance of the specific CEF level scheme and the
presence of the Fe SDW for the large magnetostriction of PrFeAsO.

\begin{figure}[tb]
\begin{center}
\includegraphics[width=0.48\textwidth]{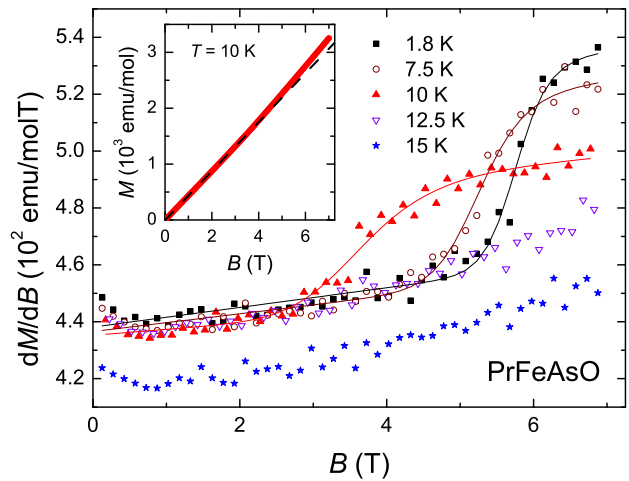}
\end{center}
\caption{The inset shows an example for the field dependence of the
magnetization $M(B)$ of PrFeAsO at 10~K. $M(B)$ exhibits an almost
linear behavior with small deviations above approximately 4~T, as
illustrated by the dashed line. They are better seen as a step in
$\mathrm{d}M/\mathrm{d}B$ shown in the main plot for different
temperatures. The solid lines are a fits to an empiric formula as
explained in the text. For clarity, not all investigated
temperatures are shown. \label{mag}}
\end{figure}

Now we turn to the transition observed in the magnetostriction
coefficient of PrFeAsO. It is expected to give also rise to an
anomaly in the magnetization $M(B)$ of the material. We performed
measurements of $M(B)$ in fields up to 7~T and for temperatures
between 1.8~K and 50~K. The magnetization is found to increase
almost linearly with field. A small change in slope is found for
$T<12.5$~K, that is more obvious from the derivative
$\mathrm{d}M/\mathrm{d}B$. As an example, the inset of
Fig.~\ref{mag} shows $M(B)$ measured in 10~K. Deviations from a
linear behavior are observed above 4~T as indicated by the dashed
line. This corresponds to a step at this field in the derivative
$\mathrm{d}M/\mathrm{d}B$ shown in the main plot of Fig.~\ref{mag}
for different temperatures. With increasing temperature, the
transition field $B_0$ shifts to lower $B$. At 12.5~K the transition
has disappeared. In order to determine the position of the
transition we fitted $\mathrm{d}M/\mathrm{d}B$ with an empiric
formula $\mathrm{d}M/\mathrm{d}B = \mathrm{A}_0 + \mathrm{A}_{1}B -
h/[1+(B/B_0)^z]$. For $z\geq 2$ this equation describes a broadened
step-like function with step hight $h$. The broadening is determined
by the parameter $z$. At $B_0$ half of the step height is reached.
The parameters A$_0$ and
A$_1$ account for a linear background. As a measure for the
uncertainty in $B_0$ we take the full width at half maximum (FWHM)
of the derivative of the fit. The fits are also shown in the main
plot of Fig.~\ref{mag}.

\begin{figure}[tb]
\begin{center}
\includegraphics[width=0.48\textwidth]{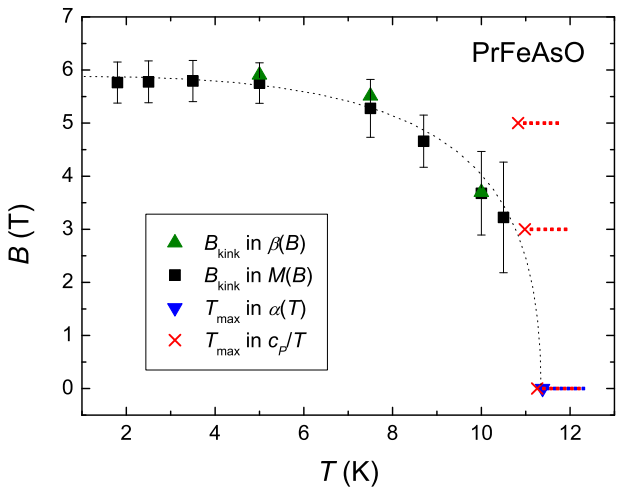}
\end{center}
\caption{The magnetic phase diagram of PrFeAsO, which was obtained
from the magnetostriction (green triangles) and the magnetization
(black squares) data. The error bars for the magnetization results
have been determined from the width of the transition in
$\mathrm{d}M/\mathrm{d}B$ as explained in the text. The red crosses
indicate the positions of the maxima in $c_p/T$, while the blue
triangle corresponds to the maximum in the thermal expansion
coefficient. The adjacent dotted lines mark the potential range for
the phase transitions. The zero-field ranges for $c_p/T$ and
$\alpha$ are almost identical and therefore hard to distinguish in
the plot. The dashed line is a guide to the eye. \label{phas}}
\end{figure}

Taking the transition fields determined from the magnetostriction
coefficient and the $B_0$ values determined from the magnetization
curves, we can draw a magnetic phase diagram for PrFeAsO, which is
shown in Fig.~\ref{phas}. The error bars given for $B_0$ are
determined from the FWHM in the derivatives of the fits as explained
above. Unfortunately, our specific heat and thermal expansion data
do not allow extracting clear transition temperatures. In
particular, no unambiguous conclusion on the character of the
transition even in zero and small magnetic fields is possible since
it can be either continuous or weakly first-order. However, the
transition is clearly suppressed in external magnetic field and the
peak maxima provide a lower limit for the transition temperatures.
The actual transitions may occur between the maxima in $c_p/T (T)$
and $\alpha (T)$ and the kinks in the data at the end of the
anomalies. These ranges are marked by dotted lines in
Fig.~\ref{phas}. In addition we also plot the peak maxima of the
anomalies.

Fig.~\ref{phas} evidences what has been noted before: The kinks in
$M(B)$ and $\beta(B)$ are observed only below $T\mathrm{_N^{Pr}}$.
In fact, the maxima of the anomalies in $c_p/T (T)$ and $\alpha (T)$
in zero field fit well to the phase line. However, the specific-heat
anomaly in 5~T and most probably also the one in 3~T is observed at
higher $T$. More precisely: the line marking the possible range for
the Pr magnetic ordering transition in 5~T clearly does not cut the
phase border. This suggests, that the transition observed in $M(B)$
and $\beta(B)$ has a different origin and is not caused directly by
the field-induced suppression of the Pr order. In fact, the
observation of a kink in both properties is a rather unusual
behavior for such an effect. Typically an abrupt reorientation of
magnetic moments shows up as step-like anomalies in $\beta (B)$ and
$M(B)$.

One may speculate that the presence of two magnetic subsystems with
competing magnetic structures is responsible for the observed
behavior. A model for the magnetic structure of PrFeAsO has been
proposed from $\mu$SR experiments:~\cite{F4-09-5} Below 127~K, the
Fe moments order in an antiferromagnetic stripe structure parallel
$a$.~\cite{F4-08-10} The resulting internal field induces a magnetic
moment on the Pr$^{3+}$ site along $c$. The Pr order below
$T_\mathrm{N}^{\mathrm{Pr}}$ leads to a reorientation of both
subsystems. The Pr$^{3+}$ moments are now oriented along $a$, which,
in turn, leads to a canting of the iron moments with a significant
component along $c$. A Fe spin reorientation upon Pr ordering was
also deduced from M\"{o}ssbauer spectroscopy.~\cite{F4-09-3}
Thus, it appears, that the
preferred magnetic structures of the Pr and Fe subsystems compete
with each other.

We suggest the following picture for PrFeAsO: In zero field the Pr
moments order below $T_\mathrm{N}^{\mathrm{Pr}}$, which is
accompanied by a canting of the Fe spins. Application of a magnetic
field destabilizes the antiferromagnetic Pr order as confirmed by
the observed small shift of the specific heat anomaly to lower $T$.
This weakening of the Pr order allows the Fe sublattice to keep its
preferred orientation down to lower $T$. Above a field of about 6~T
the Pr ordering is suppressed completely and only a polarization by
the Fe SDW and the external field remains. With regard to our
field-dependent measurements, the weakening of the Pr order leads to
a decoupling of the Fe SDW above a certain field, at which the Fe
spins resume their preferred orientation. The kink observed in our
magnetostriction data then does not correspond to a sudden flip of
Pr moments upon increasing field, but rather to a slow rotation,
which changes speed, as the Fe SDW is released. This picture of a
field-induced rotation of the Pr moments accompanied by a decoupling
of the Fe SDW is also supported by recent measurements of the
resistivity $\rho$.~\cite{F4-11-2} A weak maximum was observed in
$\rho(T)$ around 6~K in fields up to 6~T. Since the the FeAs layers
are responsible for the charge transport it was attributed to a
reorientation of the Fe moments induced by the Pr order below
$T\mathrm{_N^{Pr}}$. The authors corroborated their interpretation
by the observation of a broad hump around 6~K in the derivative of
the specific heat divided by temperature. This latter feature may be
also ascribed to the Schottky anomaly from the excited doublet at
41~K, which has a maximum in d$(c_\mathrm{mag}/T)$/d$T$ at 6.7~K. On
the other hand, the interpretation of the resistivity maximum as a
signal of Fe moment reorientation caused by the Pr order is in line
with our discussion: The maximum in $\rho(T)$ is observed only in
fields below 6~T. At higher fields the Pr order is very weak or even
destroyed as confirmed by the tiny anomaly in $c_p$ observed in 7~T
(see Fig.~\ref{cPvsT}b). As a result the Fe moments keep their
preferred orientation and the maximum in $\rho(T)$ disappears.
Moreover, a kink was found in $\rho(B)$ at low temperatures by
several groups.~\cite{F4-11-3,F4-11-2,F4-12-1} The positions of
these kinks fit rather well to our phase line, whereas small
deviations may be due to the slightly different $T\mathrm{_N^{Pr}}$
values. Therefore, we suggest that these kinks are also a signature
of the field-induced rotation of the Pr ions accompanied by Fe spin
reorientation.


\subsection{Summary }

In summary we have studied the Pr magnetism in PrFeAsO$_{1-x}$F$_x$
($x = 0, 0.15$). The CEF level scheme of the Pr$^{3+}$ ions is
strongly influenced by fluorine doping. The first excited CEF level
in PrFeAsO is found around 40~K, whereas a significantly lower
splitting corresponding to about 23-26~K is determined for
PrFeAsO$_{0.85}$F$_{0.15}$. This lowering as well as a different CEF
splitting for Pr sites with different environment may be the reason
for the absence of Pr order in the fluorine-doped compound.

At low temperatures, a field-induced transition is found for
PrFeAsO, that is not directly related to the suppression of the AFM
Pr order. Instead it comes from the interplay between the Fe and Pr
moments, which appears to be very sensitive to application of
magnetic fields. The transition is attributed to a reversal of the
Fe moment canting induced by the Pr ordering as the Pr order is
suppressed in magnetic field.

\begin{acknowledgments}
We thank M. Deutschmann, S. M\"uller-Litvanyi, R.~M\"uller, J.
Werner, and S. Ga{\ss} for technical support. Funding by the
Deutsche Forschungsgemeinschaft (DFG) within the Priority Programme
1458 (grant no. BE1749/13) is gratefully acknowledged. SW
acknowledges funding by DFG in project
WU595/3-1.\end{acknowledgments}

\bibliographystyle{PRBstyle}
\bibliography{PrFeAsO_TE}

\end{document}